\documentclass[a4paper,fleqn,usenatbib]{mnras}

\usepackage{newtxtext,newtxmath}

\usepackage[T1]{fontenc}
\usepackage{ae,aecompl}

\usepackage{graphicx}	
\usepackage{amsmath}	
\usepackage{amssymb}	

\title[Candidates for non-pulsating stars in the LMC]{Candidates for non-pulsating stars located in
the Cepheid instability strip in the Large Magellanic Cloud based on Str\"omgren photometry}

\author[W. Narloch et al.]{
Weronika Narloch,$^{1,2,3}$\thanks{E-mail: wnarloch@astro-udec.cl (WN)}
G. Pietrzy\'nski,$^{1,3}$
Z. Ko\l{}aczkowski,$^{3}$
R. Smolec,$^{3}$
M. G\'orski,$^{1,2}$
\newauthor M. Kubiak,$^{5}$
A. Udalski,$^{5}$
I. Soszy\'nski,$^{5}$
D. Graczyk,$^{1,2,4}$
W. Gieren,$^{1,2}$
P. Karczmarek,$^{5}$
\newauthor B. Zgirski,$^{3}$
P. Wielg\'orski,$^{3}$
K. Suchomska,$^{5}$
B. Pilecki,$^{3}$
M. Taormina,$^{3}$
M. Ka\l{}uszy\'nski$^{3}$
\\
$^{1}$Universidad de Concepci\'on, Departamento de Astronomia, Casilla 160-C, Concepci\'on, Chile\\
$^{2}$Millennium Institute of Astrophysics, Santiago, Chile\\
$^{3}$Nicolaus Copernicus Astronomical Center, Polish Academy of Sciences, Bartycka 18, 00-716 Warsaw, Poland\\
$^{4}$Nicolaus Copernicus Astronomical Center, Polish Academy of Sciences, Rabia\'nska 8, 87 - 100 Toru\'n, Poland\\
$^{5}$Warsaw University Observatory, Al. Ujazdowskie 4, 00-478 Warsaw, Poland
}

\date{Accepted XXX. Received YYY; in original form ZZZ}

\pubyear{2015}
\begin{document}
\label{firstpage}
\pagerange{\pageref{firstpage}--\pageref{lastpage}}
\maketitle

\begin{abstract}
We present candidates for non-pulsating stars lying in the classical Cepheid instability strip
based on OGLE photometric maps combined with Str\"omgren photometry obtained with the $4.1$-m SOAR
telescope, and Gaia DR2 data in four fields in the Large Magellanic Cloud. We selected $19$
candidates in total. After analysis of their light curves from OGLE surveys we found that all these
stars appear to be photometrically stable at the level of a~few mmag. Our results show that
non-pulsating stars might constitute to about $21\%-30\%$ of the whole sample of giant stars located
in the classical instability strip. Furthermore, we identified potential candidates for classical
Cepheids with hot companions based on their Str\"omgren colours.
\end{abstract}




\section{Introduction}

The instability strip (IS) is a~region on the Herzsprung-Russell (HR) diagram occupied by different
classes of pulsating variables \citep{Cox1974,GS1996}, but a~number of early photometric studies showed
that significant fraction of stars lying in the Cepheid region of the IS are photometrically stable at
the level of tens of mmag \citep[e.g.][]{fh1971,Schmidt1972,Percy1979}. \citet{Butler1998} studied 15
such stars lying in or close to IS of Cepheids using photometry and spectroscopy, and showed that most
of them indicate, in fact, a~variety of behavior. Only four stars did not show any variability on
a~specified level, for rest of them periodograms of Doppler velocities demonstrated multiple peaks.
The author argued that the observed variability might be caused by orbiting planetary mass companions,
rotational modulation of surface features or nonradial pulsations. He also reported larger chromospheric
activity in the non-Cepheids, but did not explain why the two classes of stars differ.

\citet{Guzik2013} used \textit{Kepler} data to examine light curves of a~sample of $633$
stars that were likely to be in or near the IS of $\delta$\,Sct and $\gamma$\,Dor stars,
among which they found $359$ \textit{stable} stars. Six of the non-pulsating objects were
lying within the IS boundaries. The authors argue that uncertainties on the temperatures
and gravities for their \textit{stable} objects may mean that some of the stars may move into
or out of the IS regions. If these uncertainties are random and not systematic then about $2\%$
of their sample would be within the IS boundaries. Later, \citet{Guzik2015} presented results
for additional $2137$ stars among which they found another $34$ such objects ($1.6\%$). As
possible explanations of the lack of pulsations the authors of these papers mention: pulsation
in different frequencies hard to detect, turning off pulsations due to various mechanisms
or simply an error in determination of $\log{g}$ or $T_{\rm eff}$. Also \citet{Murphy2015}
found stars in the $\delta$\,Sct IS that do not pulsate in p modes at the 50-$\mu$mag limit,
using \textit{Kepler} data. The authors investigated the possibility that the non-pulsators
inside the IS could be unresolved binary systems, having components that both lie outside
the IS. That interpretation would explain most of the analyzed systems, except one star which
resided in the center of the IS.

\citet{Gieren2015} reported a classical Cepheid in an eclipsing binary system OGLE-LMC-CEP-4506,
in which the secondary component turned out to be a non-pulsating red giant residing in the
center of the classical IS, which was later confirmed by \citet{Pilecki2018}. Both stars of
this system have similar masses, radii and colours, with the Cepheid being more evolutionary
advanced. Within uncertainties both stars have the same effective temperature ($T2/T1 = 0.99$),
and only slightly different luminosities ($L2/L1 = 0.83$). The orbit of the binary is highly
eccentric and the orbital period is long -- about 4.2 years. According to the results of the
analysis of six similar systems presented in \citet{Pilecki2018}, all other six stars residing
in the IS pulsate (including a~system composed of two Cepheids) and there is no pulsating star
outside the IS.

Furthermore, recently \citet{Rozyczka2018} confirmed that variable V4 in the globular cluster
M10 (NGC6254) suspected to be RR Lyr-type star \citep[][2017 edition]{Clement2001}\footnote{Webpage:
\url{http://www.astro.utoronto.ca/~cclement/cat/C1654m040}} is in fact constant.
Still, on the base of its proper motion (PM) and distance from the cluster center, the star
is a~member of M10, so its position on the colour-magnitude diagram (CMD), inside the IS, is
not a~coincidence.

These few examples show that non-pulsating stars in the IS not only exist but are in fact not
so rare. That raises the questions: how many such stars are in the IS? Is it a~significant number
or just isolated cases? In this work we try to find the answers to these questions using for our
purposes images of the Large Magellanic Cloud (LMC) taken in Str\"omgren filters. The Str\"omgren
photometric system ($uvby$) is a~four-colour medium-band photometric system (plus H$\beta$ filters)
used for stellar classification \citep{Crawford1987}. The $y$ filter is well correlated with the
$V$-band from Johnson-Cousins filters; colour $(b-y)$ determines well temperatures of stars, as well
as interstellar reddening; index $m1=(v-b)-(b-y)$ is sensitive to stellar metallicity and
$c1=(u-v)-(v-b)$ to the surface gravity. All these features make this photometric system useful when
it comes to selection of stars of a~given type.


The next important questions to answer in the folllow-up work will be: what is the nature of these
objects? What might be possible causes of the lack of pulsations in these stars? One of the potential
reasons worth to consider is the existence of strange modes (large l number) which would not cause the
changes in light curve but would affect the spectra lines of stars by broadening them. This is still
a~challenge for the theory of pulsations which does not provide a clear explanation for such situation.
Observational premises for this phenomenon would be very valuable to supplement and develop theoretical
knowledge.

The paper is organized as follows: in Section~\ref{sec:obs} we describe our data and
reduction procedure, in Section~\ref{sec:selcan} we present the method of selecting
candidates for non-pulsating stars lying in the IS, Section~\ref{sec:comm} summarizes
our search for variability in the OGLE light curves of the selected candidates and provide
a~short discussion on our search,
and finally Section~\ref{sec:concl} contains a~brief summary.

\section{Observations and data reduction}
\label{sec:obs}

Single images of four fields in the LMC (numbered 1, 2, 3 and 4) analyzed in this paper were
collected within the ARAUCARIA project \citep{Gieren2005} during one night on December $17^{th}$
2008 (programme ID: SO2008B-0917, PI: Pietrzy\'nski). The fields were chosen so that they contained
a~relatively large amount of Cepheids. Observations in four Str\"omgren filters $u$, $v$, $b$ and
$y$ were obtained using the $4.1$-m Southern Astrophysical Research Telescope (SOAR) on Cerro
Pach\'on in Chile, equipped with SOAR Optical Imager (SOI). SOI is an imager using a~mini-mosaic
of two E2V $2k \times 4k$ CCDs with a~total field of view covering $5.26 \times 5.26$~arcmin$^2$
at a~pixel scale of $0.077$~arcsec/pixel. We used pixel binning which resulted in a~pixel scale
of $0.154$~arcsec/pixel. Each chip was read out by two amplifiers. Table~\ref{tab:radec}
summarizes RA and DEC of the analyzed fields. The average seeing for each filter was
$0.81$~arcsec for $y$ and $b$, $0.89$~arcsec for $v$ and $0.91$~arcsec for $u$.

\begin{table}
	\centering
	\caption{Information about analyzed fields. Field~- number of the field in the LMC;
	RA, DEC~- equatorial coordinates of the center of the fields; E(B-V)~- calculated
	reddening value.}
	\label{tab:radec}
	\begin{tabular}{cccc}
		\hline
		Field & RA & DEC & E(B-V)\\
		 & J2000.0 & J2000.0 & [mag]\\
		\hline
		$1$ & $05\rm{:}17\rm{:}07.850$ & $-69\rm{:}21\rm{:}35.500$ & $0.1114$\\
		$2$ & $05\rm{:}18\rm{:}02.196$ & $-69\rm{:}43\rm{:}35.904$ & $0.08795$\\
		$3$ & $05\rm{:}18\rm{:}17.726$ & $-69\rm{:}36\rm{:}57.215$ & $0.08590$\\
		$4$ & $05\rm{:}28\rm{:}50.495$ & $-69\rm{:}51\rm{:}44.017$ & $0.09976$\\
		\hline
	\end{tabular}
\end{table}

Images were calibrated with standard bias subtraction and flatfield correction for each amplifier
separately. In the next step, profile photometry was performed with standard DAOPHOT/ALLSTAR package
\citep{Stetson1987} assuming a~Gaussian function with spatial variability to define point spread function
(PSF). Additionally, to reduce the effect of PSF variability, images from each amplifier were divided
into smaller, overlapping subframes. Master list for each subframe was obtained iteratively, gradually
decreasing the detection threshold. In the last iteration images were examined by eye to add manually
stars omitted in the automatic procedure. In the end, aperture corrections, calculated with DAOGROW
package \citep{Stetson1990}, were applied to each subframe and instrumental CMDs were constructed.
The average errors of our photometry were $0.02$~mag in $(b-y)$, $0.03$~mag in $m1$ and $0.04$~mag
in $c1$ for stars with brightness $V<19$~mag.

\begin{figure}
	\includegraphics[width=\columnwidth]{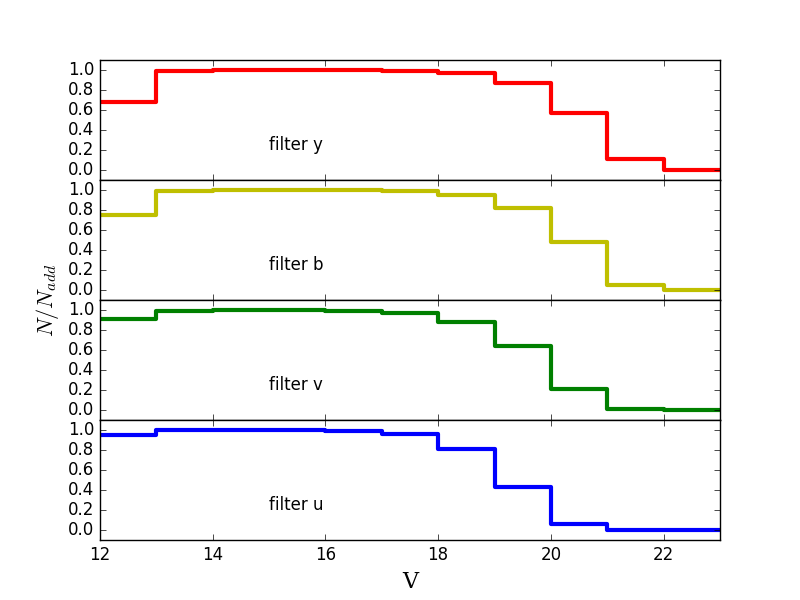}
    \caption{Completeness of our photometry in four Str\"omgren filters.}
    \label{fig:compl}
\end{figure}

The completeness of our photometry was checked with DAOPHOT package by adding one hundred artificial
stars to each subframe in filter $y$. Twenty such images were produced in each subfield of each field.
Next, the same set of stars was added to the images in the other three filters, so that in every filter
were checked the retrieving rates of the same stars. Statistics in all four fields are similar. For stars
with brightness $13<V<18$~mag the completeness is at the level of about $100\%$ in all four filters.
Stars brighter than $13th$ magnitude are often overexposed which results in decrease of completeness to
the level of about $70\%$ but no lower than $56\%$ in all filters. Completeness of stars from magnitude
bin of $(18,19)$ is still over $90\%$ in $y$, $b$ and $v$ but slighlty lower in $u$. In the range
$(19,20)$ it decreases to about $87\%$ in $y$, $82\%$ in $b$, $64\%$ in $v$ and only $43\%$ in $u$.
Completeness for stars from magnitude range $(20,21)$ drops significantly to about $57\%$ in $y$, $48\%$
in $b$, $21\%$ in $v$ and only about $6\%$ in $u$. In a~bin $(21,22)$, completeness is only about $10\%$
and $5\%$ for filters $y$ and $b$, respectively. For stars fainter than $22$ magnitude it is practically
zero in all filters. Figure \ref{fig:compl} presents completeness in four Str\"omgren filters.

\begin{figure}
	\includegraphics[width=\columnwidth]{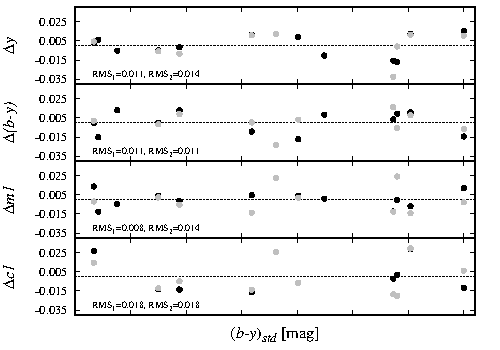}
    \caption{Residua and \textit{rms} of photometric calibration to standard system. Black and grey
    points mark residua for chip $1$ and $2$, respectively.}
    \label{fig:stdres}
\end{figure}

Magnitude $y$, colour $(b-y)$ and indices $c1$ and $m1$ were standarized for each chip individually
based on standard stars observed during the same photometric night. From $14$ stars observed in $10$
fields on chip $1$, $12$ stars were used for photometric calibration of $y$, $(b-y)$ and $m1$ and $8$ to
calibrate $c1$. From $10$ stars observed in $10$ fields on chip $2$, $9$ stars were used for photometric
calibration of $y$ and all $10$ to calibrate $(b-y)$, $m1$ and $c1$. Standard values were taken from
the \citet{Paunzen2015} catalog. The following transformations, normalized to $1$ second, were used for
chip $1$:

\begin{align}
    y_i &= (1.185 \pm 0.033) + (0.996 \pm 0.003)\cdot V \notag\\
    &- (0.024 \pm 0.019)\cdot (b-y)_{s} + (0.104 \pm 0.012)\cdot X'\\
    (b-y)_i &= (0.059 \pm 0.007) + (0.973 \pm 0.016)\cdot (b-y)_{s} \notag\\
    &+ (0.066 \pm 0.012)\cdot X'\\
    m1_i &= (0.327 \pm 0.006) + (0.904 \pm 0.027)\cdot m1_{s} \notag\\
    &+ (0.189 \pm 0.024)\cdot (b-y)_{s} + (0.051 \pm 0.010)\cdot X'\\
    c1_i &= -(0.558 \pm 0.073) + (1.139 \pm 0.078)\cdot c1_{s} \notag\\
    &+ (0.014 \pm 0.087)\cdot (b-y)_{s} + (0.142 \pm 0.024)\cdot X' \\ \notag
	\label{eq:standarization_s1}
\end{align}
and chip $2$:

\begin{align}
    y_i &= (1.176 \pm 0.216) + (0.998 \pm 0.021)\cdot V \notag\\
    &- (0.002 \pm 0.037)\cdot (b-y)_{s} + (0.113 \pm 0.020)\cdot X'\\
    (b-y)_i &= (0.051 \pm 0.009) + (0.973 \pm 0.019)\cdot (b-y)_{s} \notag\\
    &+ (0.054 \pm 0.012)\cdot X'\\
    m1_i &= (0.322 \pm 0.012) + (0.929 \pm 0.057) \cdot m1_{s} \notag\\
    &+ (0.128 \pm 0.053)\cdot (b-y)_{s} + (0.065 \pm 0.017)\cdot X'\\
    c1_i &= -(0.641 \pm 0.052) + (1.116 \pm 0.058) \cdot c1_{s} \notag\\
    &+ (0.018 \pm 0.067)\cdot (b-y)_{s} + (0.136 \pm 0.022)\cdot X',
	\label{eq:standarization_s2}
\end{align}
where $X'=X-1.25$ is an airmass; $y_i$, $(b-y)_i$, $m1_i$ and $c1_i$ are instrumental and $V$,
$(b-y)_{s}$, $m1_{s}$ and $c1_{s}$ are standard magnitudes and indices, respectively.
Figure \ref{fig:stdres} presents residua and \textit{rms} of applied transformations for both chips.
To check the internal consistency of our transformations between chips we fitted the Paczy\'nski
profile \citep{ps1998} to the histogram of red clump (RC) from chip $1$ and $2$, and found the center
of the pick of the distribution. In the equation \ref{eq:RC} of the following form \citep{Gorski2018}:

\begin{align}
	n(k) &= a + b(k-k_{RC}) +c(k-k_{RC})^2 \notag\\
	&+ \frac{N_{RC}}{\sigma_{RC}\sqrt{2\pi}}exp\left(-\frac{(k-k_{RC})^2}{2\sigma_{RC}}\right)
	\label{eq:RC}
\end{align}

the Gaussian component represents a~fit to the RC itself and two other terms describe the background
distribution of the red giant stars. $k$ state for the colour $(b-y)$, $k_{RC}$ is the mean colour
of the RC stars, $\sigma_{RC}$ the spread of the RC stars colour and $N_{RC}$ a~number of RC stars.
The shift in colour between RC from both chips in all four fields was less than $0.02$~mag, which is
an acceptable difference.

Our final list of stars contains $26291$ objects measured to a limiting magnitude of about $21.5$~mag
in all four filters simultaneously.

\section{Selection of candidates}
\label{sec:selcan}

\begin{figure}
	\includegraphics[width=\columnwidth]{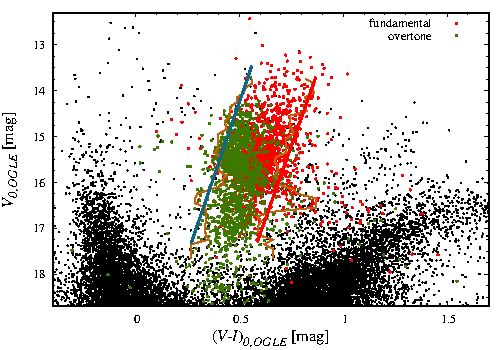}
    \caption{Dereddened OGLE CMD for stars brighter than $19$~mag from our four fields in the LMC.
    Red and green points mark fundamental and overtone Cepheids from OGLE Collection of Variable Stars,
    respectively. Orange curves show the result of applying step detection technique. Strait blue and
    red lines are fitted to data from previous step and represent adopted borders of empirical IS for
    classical Cepheids from LMC.}
    \label{fig:cmdisOGLE}
\end{figure}

In the first step of the selection of candidates for non-pulsating stars residing in the classical
IS of Cepheids in the LMC we determined empirical boundaries of the IS. For this we used the
photometry of over $3200$ Cepheids from the LMC gathered in the OGLE Collection of Variable Stars
and having measurments in both $V$ and $I$ filters. The position of stars in the CMD is affected by
reddening caused by the interstellar extinction along the line of sight. To correct for this effect
we used the reddenings calculated for each field with Cepheids based on the colour of the red clump
\citep[][G\'orski et al. 2019, submitted]{Gorski2018}. Its true colour was determined in $20$ fields
containing late type eclipsing binaries with reddening values calculated in \citet{Graczyk2018} based
on calibration between the equivalent width of the interstellar absorption Na~I~D1 line and the
reddening. The uncertainties of the reddening values in a~given field consist of the adopted true
colour of the red clump and statistical error of apparent colour measurments where the latter is
dominant and are $\sigma E(B-V)=0.022$~mag. The details of this calculation will be presented in
G\'orski et al. (2019, submitted). Next, in the overlapping magnitude bins of $0.5$~mag wide in $V$
shifted downwards by $0.1$~mag in the range $(13.5-17.5)$~mag we
calculated histograms of colours and then used step detection technique to find the empirical blue and
red edges of the instability strip. Later, we fitted a~strait line to these points. This way, we got
rather conservative empirical position of the instability strip of Cepheids in the LMC calculated based
on thousands of Cepheids (see Figure~\ref{fig:cmdisOGLE}). Stars that are located within IS, but
apparently are non-variable, will be our candidates for non-pulsating stars within classical IS.

In the next step, we downloaded the BVI OGLE photometry \citep{Udalski2000} for our four fields and
identified the Cepheids to later exclude them from the sample of potential candidates for
non-pulsating stars. We identified $42$ classical Cepheids \citep{Soszynski2015} and two type-II
Cepheids \citep{Soszynski2018}, which gave $44$ stars in total, summarized in Table~\ref{tab:cepheids}
($11$, $10$, $10$ and $13$ Cepheids in the fields $1$, $2$, $3$ and $4$, respectively) where we also
give their Str\"omgren photometry. From the VI CMD of remaining stars we selected these lying within
empirical boundaries calculated in the previous step and being no fainter than the faintest Cepheid
($V_0=17.46$~mag). At this stage, we selected $61$ stars in total.

\begin{figure}
	\includegraphics[width=\columnwidth]{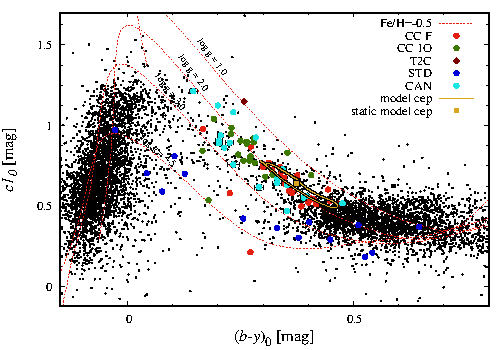}
    \caption{Dereddened Str\"omgren two-colour diagram for stars brighter than $19$~mag from our four
    fields in the LMC. Red and green dots~- stars identified in OGLE catalogs as classical Cepheids (CC)
    pulsating in F and 1O modes, respectively; dark red diamonds~- type~II Cepheids (T2C); blue dots~-
    standard main sequence stars used for standardization (STD); cyan dots~- stars selected from IS as
    candidates for non-pulsating stars (CAN); red lines~- theoretical relations for metallicity
    $\rm [Fe/H]=-0.5$~dex; yellow line~- model of Cepheid; yellow square~- static model of Cepheid.}
    \label{fig:c1byisOGLE}
\end{figure}

As next, we cross-matched the BVI OGLE photometry of candidates with our Str\"omgren photometry and
plotted two-colour diagram $c1-(b-y)$ for identified Cepheids and selected stars
(see Fig.~\ref{fig:c1byisOGLE}). Our Str\"omgren data were dereddened using relations of relative
extinctions from \citet{Schlegel1998} and the reddening values in our four fields which are given in
Table~\ref{tab:radec}. On the same diagram we plotted theoretical $c1-(b-y)$ relations for
stars from different luminosity classes (having different surface gravity expressed by $\log{g}$) from
static atmosphere models of \citet{ck} for metallicity $\rm [Fe/H]=-0.5$ (closest to average metallicity
of Cepheids in the LMC). The periods for most of identified classical Cepheids range between
$\approx 0.65$~d to $\approx11.2$~d which correspond to $\log{g} \approx 2.8$ for F mode or $2.7$ for
1O mode and $\log{g} \approx 1.5$ for F mode or $\approx 1.3$ for 1O mode, respectively, where most
stars have $\log{g} \approx 2.0$ (R. Smolec, priv. comm.). So we shifted theoretical $c1-(b-y)$
relations accordingly in $(b-y)$ colour by $-0.05$~mag to match our observations. We then decided to
select stars lying between theoretical lines for $\log{g} = 1.0$ and $\log{g} = 3.0$ to cover wide range
of possible values of $\log{g}$ of Cepheids. This way, we selected giant stars from the sample of
potential candidates for non-pulsating stars obtained in the previous step and reduced their number to
$27$ stars. This procedure is presented in Fig.~\ref{fig:c1byisOGLE} where we additionally added
a~theoretical line for $\log{g} = 4.5$ (close to solar value) and marked our dereddened standard main
sequence stars. They arrange systematically below Cepheids which proves the potential of $c1$
Str\"omgren index in separating stars of different luminosity classes. During the pulsation cycle, the
effective temperature and brightness of a~Cepheid changes and hence the star performs a~loop either in
the CMD or $c1-(b-y)$ relation. It may even leave the instability strip during some pulsation phases.
To illustrate this effect, and to estimate how large the excursion beyond the adopted boundaries can be,
we have computed a~non-linear Cepheid model with $M=4.5\,{\rm M}_{\odot}$, $L=1784\,{\rm L}_{\odot}$,
${\rm [Fe/H]}=-0.3$~dex (where convective parameters were a~set B adopted by \citet{Baranowski2009},
see their tab. 3). The model pulsates in the radial fundamental mode with a~period of $5.93$\,d. The
instantaneous effective temperature and gravity (including the dynamic acceleration) were used to
compute $(b-y)_0$ and $c1_0$ using the static atmosphere models of \cite{ck}. Obviously, using static
model atmospheres is not the best approach here, but we just want to estimate the typical extent of the
loop followed in the $c1-(b-y)$ relation during the pulsation. The result is plotted in
Fig.~\ref{fig:c1byisOGLE} (yellow line). Symbol (yellow square) inside the loop corresponds to the
parameters of the equilibrium model.

\begin{figure}
	\includegraphics[width=\columnwidth]{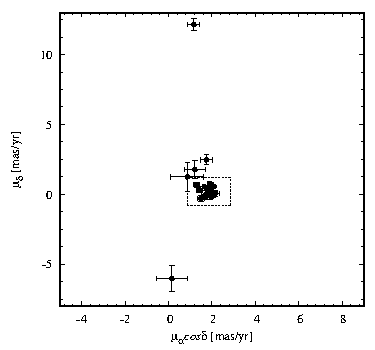}
    \caption{Vector Point Diagram for stars selected as candidates for non-pulsating stars in the
    classical IS based on Gaia DR2 catalog. Black, dashed box encloses stars with proper motions that
    coincide with the proper motion of the LMC \citep{Marel2016}, and thus belong to this galaxy.}
    \label{fig:gaia_dr2}
\end{figure}

As a~separate test, we downloaded parallaxes and PMs for selected candidates from Gaia DR2 catalog
\citep{gaia2016,gaia2018}. Fig.~\ref{fig:gaia_dr2} presents the Vector Point Diagram (VPD) for our
candidates. One star with high proper motion is not shown in the diagram. From the stars we selected
those with PMs within a~box of $\pm 1$~mas/yr around the PM of the LMC given by \citet{Marel2016}
(marked with black dashed line in Fig.~\ref{fig:gaia_dr2}). In the end, we were left with $19$
candidates in total ($3$ in field $1$, $4$ in field $2$, $7$ in field $3$ and $5$ in field $4$).
Remaining stars are most probably giant stars from our galaxy.
Table~\ref{tab:candidatesOGLE} summarizes selected objects and Table~\ref{tab:candidates} gives their
identification numbers in the OGLE catalogues. 

\begin{figure}
	\includegraphics[width=\columnwidth]{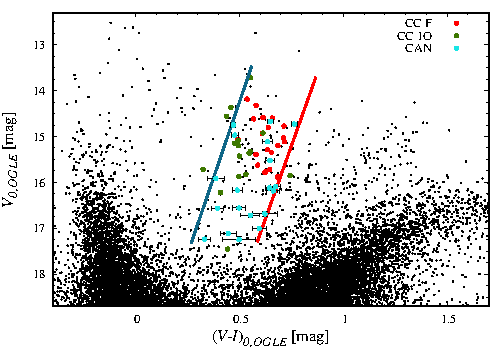}
    \caption{Dereddened OGLE CMD with marked Cepheids and candidates for non-pulsating stars in the IS.
    Red and green dots mark identified classical Cepheids (CC) pulsating in F and 1O modes, respectively.
    Cyan dots mark candidates for non-pulsating stars (CAN). The blue and red lines present empirical
    edges of IS.}
    \label{fig:cmdisOGLEcan}
\end{figure}

\begin{figure}
	\includegraphics[width=\columnwidth]{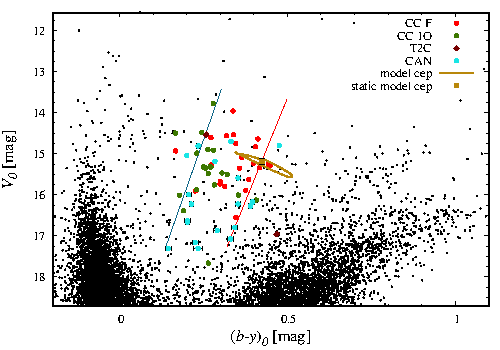}
    \caption{Dereddened Str\"omgren CMD for stars brighter than $19$th magnitude. Red and green dots
    mark stars identified as classical Cepheids (CC) pulsating in F and 1O modes, respectively. Dark
    red diamonds mark type~II Cepheids (T2C). Green dots mark candidates for non-pulsating stars (CAN).
    Blue and red lines mark the approximate position of the empirical IS determined on the basis
    of the OGLE cefeid and tranformed into Str\"omgren CMD.}
    \label{fig:cmdisSOARcan}
\end{figure}

The result of our selection is presented in Fig.~\ref{fig:cmdisOGLEcan}, where are marked candidates
for non-pulsating stars as well as identified Cepheids. On the other hand, Fig.~\ref{fig:cmdisSOARcan}
shows the candidates and Cepheids in the Str\"omgren CMD which is an independent check of the reliablity
of our selection procedure. Candidates selected from OGLE CMD should fall into IS also in Str\"omgren
CMD. To check if that is the case, we tranformed the IS boundaries from OGLE CMD into Str\"omgren CMD
using 2nd order polynomial of the form
$(b-y)_0 = 0.0965464 (V-I)_0^2 + 0.493749 (V-I)_0 - 0.0028315$ (marked in Fig.~\ref{fig:cmdisSOARcan}
with dashed blue and red lines). Most of the candidates fit very well into IS region showing that
either selection from the OGLE CMD or Str\"omgren CMD would classify them as non-pulsating stars in
the classical IS. However, few stars lie beyond the IS egdes. This might be the effect of approximated
transformation of the IS or possible blending of candidates in either OGLE or Str\"omgren CMD.
Also, most of identified Cepheids fit very well in the approximated IS, except few, which however is not
surprising, since their position on the Str\"omgren CMD is based on a~single colour measurement.
As it was already mentioned in previous paragraph, the Cepheids performe a~loop in CMD when they change
their effective temperature and brightness during its pulsation cycle and in particular phases can
go beyond the IS boundaries. In Fig.~\ref{fig:cmdisSOARcan} we marked the results of a~non-linear
Cepheid model from the $c1_0-(b-y)_0$ relation. The horizontal extent of the loop in the CMD exceeds
$0.15$\,mag in $(b-y)_0$. The most profunding Cepheids probably are simply catched in a~specific phase.
The only exception seems star OGLE-LMC-T2CEP-086 from field $2$, which is type~II Cepheid located rather
well beyond the classical IS. Type~II Cepheids are characterized by different physical parameters
however; in particular, they have significantly lower masses, akin to that of RR~Lyrae stars. This
leads to a~wider instability strip at lower luminosities, see \cite{s16}.

%
\section{Selected candidates}
\label{sec:comm}

We selected $19$ candidates for non-pulsating stars lying in the IS based on the OGLE CMD combined
with the Str\"omgren photometry which is about $30\%$ of the whole sample of Cepheids and non-pulsating
objects located in the classical IS in the LMC. The confrontation of this result with Str\"omgren CMD
put seven of them under the question which would leave us with about $21\%$.
We analyzed the light curves of the candidates downloaded from OGLE-II database
\citep{Szymanski2005}\footnote{Available at: \url{http://ogledb.astrouw.edu.pl/~ogle/photdb/}} and
nonpublic catalogs of OGLE-III and OGLE-IV surveys to check if we might be able to detect any signs
of variability. During analysis we arbitrarily assumed signal-to-noise of $4$ as a~threshold for
reliable detection of significant periodicity in the frequency spectrum.
For all selected candidates no significant variability was found at the level of a~few mmag,
so we can assume that these stars are constant at this level of detection.

We shall consider the possibility that our candidates fall into IS because of the blending caused
by crowding or because they are physically associated with other stars and create unresolved
binaries or multiple systems, which would change their magnitudes and colours. The second case
was discussed in detail by, e.g., \citet{Mochejska2000} where the authors showed that this effect
is significant in case of Cepheids. For this example, \citet{Pilecki2018} analyzed Cepheids in
eclipsing binaries in the LMC. They studied seven such systems where one was composed of two classical
Cepheids, four were binaries composed of classical Cepheid and constant star and one with a~type~II
Cepheid and constant star. The authors derived precise physical stellar parameters for all components
of the systems. This is too small sample of stars to conclude any typical features of binaries
with Cepheids but it gives us a~concept of how big may be the change of colour of a~Cepheid with
a~companion star. The smallest $(V-I)$ colour difference of a~single classical Cepheids relative to the
colour of the binary was $0.005$~mag and the biggest $0.19$~mag which correspond to $\approx 0.003$ and
$\approx 0.1$~mag,  respectively, in Str\"omgren system.  The exceptionally large change of colour had
a~system with type~II Cepheid which was about $0.7$~mag ($\Delta (b-y) \approx 0.4$~mag). Interestingly,
in all systems no third light was detected. Thus, if the colour difference of a~single Cepheids and
their binaries can be so significant that they can fall out from IS then also the opposite situation
can take place when a~star originally lying outside IS could fall into it. In the extreme case, all
of our candidates might be blends. Unfortunately, we need the spectroscopy in order to investigate the
nature of the candidates. At this stage, we are not able to resolve this matter. However, we can
estimate the effect of crowding on our selected stars based on our Str\"omgren data.

In order to estimate what might be the influence of the crowding on our candidates, we performed a~series
of simulations. In each subfield of our four fields we take a~catalog with Str\"omgren photometry and
draw a~subsample of $100$ stars with magnitudes measured in all four filters simultaneously. Next,
randomly we draw x and y coordinates in pixels from a~uniform distribution from the range of the image
size. In DAOPHOT package we use task \textit{add} to add created lists of stars in a~specific filter to
the image in this filter. We repeat this procedure $100$ times for each subfield of a~given field. Then,
we performe the photometry on images with added stars and calculate the resulting colours. Subsequently,
we calculate the diffrences between the actual colours and new colours of added stars. From the latter
we can estimate what is the influence of the crowding effect in all our fields. The four considered
fields are located in the bar of the LMC and have similar star density. This is reflected in statistics
of crowding effect for all fields, which is similar. On average $53\%$ of added stars from the vincinity
of the IS changed their $(b-y)$ colour by more than $0.01$~mag, $11\%$ by more than $0.05$~mag and only
$5\%$ by more than $0.1$~mag.
Most of the candidates reside relatively deep in the IS. They would have to change their colours
significantly to enter IS as a~blend and this is very unlikely. On the other hand, there is more than
$50\%$ chance that candidates lying close the IS edges falled into it as a~consequence of blending due
to crowding. There are seven such stars in OGLE CMD. Comparison of two CMDs (OGLE and Str\"omgren) also
shows that over one third of stars could be blended due to crowding and because of that entered IS.
That still leave us with $12$ candidates for non-pulsating stars which account for $21.4\%$ of the whole
sample of giant stars in the IS region.

\subsection{Cepheids in analyzed fields}
\label{ssec:cep}

\begin{figure}
	\includegraphics[width=\columnwidth]{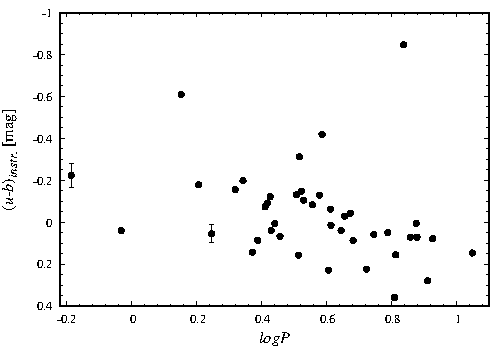}
    \caption{$(u-b) - \log{P}$ relation for classical Cepheids identified in the four fields in the LMC.}
    \label{fig:ublogP}
\end{figure}

In Fig.~\ref{fig:ublogP} we plotted relation $(u-b) - \log{P}$ for $41$ classical Cepheids identified
in our fields \citep{Soszynski2015}. Most of the stars lie in the bottom of the figure creating
a~linear relation, scatter of which is a~consequence of having only one random measurement of the
brightness. Two stars which have the largest colour differences clearly stand out from that relation.
Two others are questionable. These are very likely binaries with hot main sequence component, bright
in $u$-band. Further spectroscopic observations would be required to confirm the binary nature of those
systems.

\begin{table*}
	\centering
	\caption{Stars identified as Cepheids in OGLE catalogs from four fields in the LMC. OGLE~-
	name of the Cepheid in OGLE catalogs (OGLE-LMC-CEP-NNNN, OGLE-LMC-T2CEP-NNN); RA, DEC~-
	equatorial coordinates in degrees; field~- name of the field in the LMC; $V$~- $V$-band
	magnitude in Johnson-Cousin system from SOAR; $(b-y)$~- Str\"omgren colour; $m1$~- $m1$
	indicator; $c1$~- $c1$ indicator; type,mode~- identified type and mode of the star (CC~-
	classical Cepheid, T2C~- type-II Cepheid).}
	\label{tab:cepheids}
	\begin{tabular}{lcccccccll} 
		\hline
		OGLE & RA & DEC & field & $V_S$ & $(b-y)$ & $m1$ & $c1$ & P & type,mode\\
		 & J2000.0 & J2000.0 & & [mag] & [mag] & [mag] & [mag] & [d] & \\
		\hline
                $1454$   & $79.16290$ & $-69.36194$ & $1$ & $16.496 \pm 0.004$ & $0.493 \pm 0.006$ &
                $0.223 \pm 0.010$ & $0.713 \pm 0.010$ & $2.8663497(53)$ & CC,1O\\ 
                $1460$   & $79.21088$ & $-69.32492$ & $1$ & $14.864 \pm 0.002$ & $0.252 \pm 0.003$ &
                $0.160 \pm 0.005$ & $0.863 \pm 0.005$ & $3.8617970(67)$ & CC,1O\\ 
                $1461$   & $79.21956$ & $-69.36780$ & $1$ & $15.995 \pm 0.004$ & $0.471 \pm 0.005$ &
                $0.307 \pm 0.008$ & $0.519 \pm 0.008$ & $4.7178165(47)$ & CC,F\\ 
                $1463$   & $79.22900$ & $-69.33071$ & $1$ & $14.905 \pm 0.002$ & $0.425 \pm 0.003$ &
                $0.262 \pm 0.005$ & $0.727 \pm 0.005$ & $7.5282566(85)$ & CC,F\\ 
                $1466$  & $79.24441$  & $-69.39396$ & $1$ & $15.456 \pm 0.003$ & $0.450 \pm 0.004$ &
                $0.332 \pm 0.007$ & $0.602 \pm 0.007$ & $6.1535625(125)$ & CC,F\\ 
                $1468$   & $79.24965$ & $-69.33508$ & $1$ & $16.163 \pm 0.004$ & $0.399 \pm 0.006$ &
                $0.209 \pm 0.010$ & $0.790 \pm 0.010$ & $3.6069822(48)$ & CC,F\\ 
                $1471$   & $79.25756$ & $-69.36755$ & $1$ & $18.034 \pm 0.018$ & $0.350 \pm 0.020$ &
                $0.120 \pm 0.038$ & $0.908 \pm 0.082$ & $0.6523943(8)$ & CC,1O/2O\\ 
                $1479$   & $79.29086$ & $-69.34830$ & $1$ & $16.246 \pm 0.004$ & $0.315 \pm 0.006$ &
                $0.192 \pm 0.009$ & $0.968 \pm 0.045$ & $1.7645509(20)$ & CC,1O\\ 
                $3832$   & $79.34182$ & $-69.33540$ & $1$ & $14.324 \pm 0.004$ & $0.424 \pm 0.006$ &
                $0.232 \pm 0.011$ & $0.762 \pm 0.016$ & $11.2127522(930)$ & CC,F\\ 
                $1490$   & $79.35307$ & $-69.34935$ & $1$ & $15.008 \pm 0.002$ & $0.498 \pm 0.004$ &
                $0.336 \pm 0.007$ & $0.562 \pm 0.007$ & $8.1702063(80)$ & CC,F\\ 
                $1491$   & $79.36054$ & $-69.33521$ & $1$ & $14.854 \pm 0.003$ & $0.331 \pm 0.004$ &
                $0.243 \pm 0.007$ & $0.840 \pm 0.008$ & $4.4205628(154)$ & CC,1O\\ 
                $1499$   & $79.41005$ & $-69.75534$ & $2$ & $15.522 \pm 0.004$ & $0.468 \pm 0.005$ &
                $0.346 \pm 0.017$ & $0.537 \pm 0.030$ & $5.5756990(40)$ & CC,F\\ 
                $1500$   & $79.41714$ & $-69.69831$ & $2$ & $15.801 \pm 0.003$ & $0.389 \pm 0.004$ &
                $0.238 \pm 0.007$ & $0.711 \pm 0.007$ & $2.6768609(36)$ & CC,1O\\ 
                $1502$   & $79.42158$ & $-69.69052$ & $2$ & $15.969 \pm 0.003$ & $0.3698\pm 0.004$ &
                $0.222 \pm 0.007$ & $0.758 \pm 0.008$ & $3.3402478(37)$ & CC,F\\ 
                $1509$   & $79.46337$ & $-69.73214$ & $2$ & $14.062 \pm 0.002$ & $0.347 \pm 0.002$ &
                $0.218 \pm 0.003$ & $0.912 \pm 0.004$ & $4.5328125(189)$ & CC,1O\\ 
                $1512$   & $79.47915$ & $-69.58195$ & $3$ & $15.559 \pm 0.004$ & $0.339 \pm 0.005$ &
                $0.214 \pm 0.008$ & $0.884 \pm 0.009$ & $4.1055073(24)$ & CC,F\\ 
                $1513$   & $79.48029$ & $-69.76290$ & $2$ & $15.117 \pm 0.003$ & $0.473 \pm 0.006$ &
                $0.335 \pm 0.012$ & $0.563 \pm 0.012$ & $8.4608625(64)$ & CC,F\\ 
                $1515$   & $79.48530$ & $-69.58140$ & $3$ & $16.189 \pm 0.020$ & $0.292 \pm 0.022$ &
                $0.310 \pm 0.031$ & $0.596 \pm 0.029$ & $3.2864900(16)$ & CC,F\\ 
                $1517$   & $79.48539$ & $-69.64815$ & $3$ & $15.636 \pm 0.004$ & $0.424 \pm 0.005$ &
                $0.358 \pm 0.008$ & $0.610 \pm 0.008$ & $4.8132991(32)$ & CC,F\\ 
                $1521$   & $79.50368$ & $-69.73644$ & $2$ & $15.277 \pm 0.003$ & $0.298 \pm 0.004$ &
                $0.169 \pm 0.005$ & $1.001 \pm 0.005$ & $3.2147769(97)$ & CC,1O\\ 
                $1522$   & $79.50522$ & $-69.62087$ & $3$ & $15.187 \pm 0.003$ & $0.329 \pm 0.004$ &
                $0.188 \pm 0.005$ & $0.909\pm 0.005$ & $3.3950599(74)$ & CC,1O\\ 
                $1528$   & $79.52725$ & $-69.65014$ & $3$ & $16.021 \pm 0.004$ & $0.365 \pm 0.005$ &
                $0.212 \pm 0.008$ & $0.776 \pm 0.008$ & $3.7905721(26)$ & CC,F\\ 
                $1532$   & $79.53612$ & $-69.75633$ & $2$ & $16.844 \pm 0.005$ & $0.415 \pm 0.006$ &
                $0.088 \pm 0.011$ & $0.691 \pm 0.012$ & $2.2056640(7)$ & CC,F\\ 
                $1533$   & $79.53685$ & $-69.76236$ & $2$ & $15.637 \pm 0.003$ & $0.324 \pm 0.004$ &
                $0.252 \pm 0.007$ & $0.797 \pm 0.007$ & $2.7616841(31)$ & CC,1O\\ 
                $1536$   & $79.54853$ & $-69.58716$ & $3$ & $16.293 \pm 0.005$ & $0.420 \pm 0.006$ &
                $0.079 \pm 0.010$ & $0.846 \pm 0.010$ & $1.6112568(68)$ & CC,1O\\ 
                $1537$   & $79.55684$ & $-69.70578$ & $2$ & $15.759 \pm 0.003$ & $0.373 \pm 0.004$ &
                $0.271 \pm 0.007$ & $0.697 \pm 0.007$ & $2.6932881(31)$ & CC,1O\\ 
                $086$    & $79.57418$ & $-69.72433$ & $2$ & $17.251 \pm 0.006$ & $0.536 \pm 0.008$ &
                $0.134 \pm 0.015$ & $0.529 \pm 0.016$ & $15.8455000(829)$ & T2C/WVir\\ 
                $1549$   & $79.64756$ & $-69.60096$ & $3$ & $16.297 \pm 0.004$ & $0.245 \pm 0.006$ &
                $0.157 \pm 0.011$ & $0.552 \pm 0.015$ & $1.4226693(14)$ & CC,1O\\ 
                $1550$   & $79.66199$ & $-69.58794$ & $3$ & $16.173 \pm 0.005$ & $0.442 \pm 0.007$ &
                $0.227 \pm 0.012$ & $0.707 \pm 0.012$ & $3.2677563(18)$ & CC,F\\ 
                $1551$   & $79.67019$ & $-69.65191$ & $3$ & $16.673 \pm 0.005$ & $0.257 \pm 0.006$ &
                $0.181 \pm 0.010$ & $1.057 \pm 0.010$ & $0.9308301(14)$ & CC,2O\\ 
                $1553$   & $79.67459$ & $-69.63975$ & $3$ & $15.543 \pm 0.003$ & $0.510 \pm 0.004$ &
                $0.307 \pm 0.008$ & $0.618 \pm 0.012$ & $6.4571732(54)$ & CC,F\\ 
                $2085$   & $82.09549$ & $-69.86013$ & $4$ & $16.091 \pm 0.004$ & $0.359 \pm 0.006$ &
                $0.190 \pm 0.010$ & $0.788 \pm 0.010$ & $2.0879263(20)$ & CC,1O\\ 
                $2086$   & $82.09638$ & $-69.86773$ & $4$ & $15.643 \pm 0.003$ & $0.347 \pm 0.004$ &
                $0.242 \pm 0.007$ & $0.801 \pm 0.006$ & $2.5793555(21)$ & CC,1O\\ 
                $2090$   & $82.12354$ & $-69.87687$ & $4$ & $14.890 \pm 0.002$ & $0.396 \pm 0.003$ &
                $0.295 \pm 0.005$ & $0.743 \pm 0.005$ & $7.2263791(63)$ & CC,F\\ 
                $2094$   & $82.13316$ & $-69.84842$ & $4$ & $15.081 \pm 0.003$ & $0.423 \pm 0.004$ &
                $0.314 \pm 0.005$ & $0.659 \pm 0.005$ & $7.5701883(155)$ & CC,F\\ 
                $2099$   & $82.14641$ & $-69.85654$ & $4$ & $14.938 \pm 0.002$ & $0.349 \pm 0.003$ &
                $0.165 \pm 0.005$ & $0.232 \pm 0.005$ & $6.8938013(128)$ & CC,F\\ 
                $2106$   & $82.18173$ & $-69.83431$ & $4$ & $15.672 \pm 0.004$ & $0.494 \pm 0.006$ &
                $0.387 \pm 0.011$ & $0.532 \pm 0.017$ & $5.2938004(73)$ & CC,F\\ 
                $2107$   & $82.18535$ & $-69.83482$ & $4$ & $15.657 \pm 0.004$ & $0.349 \pm 0.006$ &
                $0.214 \pm 0.010$ & $0.826 \pm 0.010$ & $2.6243078(37)$ & CC,1O\\ 
                $2109$   & $82.18951$ & $-69.86858$ & $4$ & $15.630 \pm 0.003$ & $0.525 \pm 0.004$ &
                $0.302 \pm 0.007$ & $0.551 \pm 0.008$ & $6.5179412(38)$ & CC,F\\ 
                $2112$   & $82.21425$ & $-69.83453$ & $4$ & $15.260 \pm 0.004$ & $0.244 \pm 0.004$ &
                $0.215 \pm 0.006$ & $0.996 \pm 0.005$ & $4.1159516(19)$ & CC,F\\ 
                $129$    & $82.22751$ & $-69.87808$ & $4$ & $14.864 \pm 0.002$ & $0.335 \pm 0.003$ &
                $0.038 \pm 0.005$ & $1.166 \pm 0.005$ & $62.5089466(237959)$ & T2C/RVTau\\ 
                $2118$   & $82.26211$ & $-69.89023$ & $4$ & $15.651 \pm 0.003$ & $0.327 \pm 0.009$ &
                $0.193 \pm 0.017$ & $0.923 \pm 0.010$ & $2.4461347(26)$ & CC,1O\\ 
                $2124$   & $82.30195$ & $-69.89078$ & $4$ & $15.247 \pm 0.003$ & $0.357 \pm 0.004$ &
                $0.223 \pm 0.007$ & $0.939 \pm 0.008$ & $4.0439894(103)$ & CC,1O\\ 
                $2130$   & $82.33005$ & $-69.83155$ & $4$ & $15.813 \pm 0.008$ & $0.341 \pm 0.009$ &
                $0.209 \pm 0.022$ & $0.918 \pm 0.037$ & $2.3572510(32)$ & CC,1O\\ 
		\hline
	\end{tabular}
\end{table*}

\begin{table*}
	\centering
	\caption{Candidates for non-pulsating stars in the IS from four fields in the LMC. ID~-
	name of the candidate; RA, DEC~- equatorial coordinates in degrees; field~- number of the
	field in the LMC; $V_O$~- OGLE-II magnitude; $(V-I)_O$~- OGLE-II colour; $V_S$~- $V$-band
	magnitude in Johnson-Cousin system from SOAR; $(b-y)$~- Str\"omgren colour; $m1$~- $m1$
	indicator; $c1$~- $c1$ indicator.}
	\label{tab:candidatesOGLE}
	\begin{tabular}{cccccccccc} 
		\hline
		ID & RA & DEC & field & $V_O$ & $(V-I)_O$ & $V_S$ & $(b-y)$ & $m1$ & $c1$\\
		 & J2000.0 & J2000.0 & & [mag] & [mag] & [mag] & [mag] & [mag] & [mag] \\
		\hline
		CAN-01 & $79.19244$ & $-69.35403$ & $1$ & $15.892 \pm 0.019$ & $0.789 \pm 0.021$ &
		$15.958 \pm 0.004$ & $0.439 \pm 0.006$ & $0.231 \pm 0.009$ & $0.653 \pm 0.008$\\
		CAN-02 & $79.23039$ & $-69.36134$ & $1$ & $17.607 \pm 0.018$ & $0.473 \pm 0.027$ &
		$17.675 \pm 0.006$ & $0.231 \pm 0.009$ & $0.104 \pm 0.016$ & $1.232 \pm 0.015$\\
		CAN-03 & $79.32142$ & $-69.38572$ & $1$ & $15.340 \pm 0.012$ & $0.619 \pm 0.017$ &
		$15.407 \pm 0.003$ & $0.286 \pm 0.004$ & $0.155 \pm 0.007$ & $1.144 \pm 0.007$\\
		CAN-04 & $79.45441$ & $-69.64956$ & $3$ & $17.004 \pm 0.040$ & $0.662 \pm 0.046$ &
		$17.160 \pm 0.007$ & $0.357 \pm 0.010$ & $0.254 \pm 0.017$ & $0.632 \pm 0.018$\\
		CAN-05 & $79.45741$ & $-69.57179$ & $3$ & $16.848 \pm 0.021$ & $0.504 \pm 0.030$ &
		$16.954 \pm 0.005$ & $0.268 \pm 0.007$ & $0.202 \pm 0.012$ & $0.919 \pm 0.012$\\
		CAN-06 & $79.46599$ & $-69.75158$ & $2$ & $16.979 \pm 0.055$ & $0.734 \pm 0.061$ &
		$17.088 \pm 0.006$ & $0.411 \pm 0.009$ & $0.035 \pm 0.017$ & $0.745 \pm 0.018$\\
		CAN-07 & $79.47437$ & $-69.73521$ & $2$ & $16.411 \pm 0.017$ & $0.756 \pm 0.019$ &
		$16.518 \pm 0.005$ & $0.421 \pm 0.007$ & $0.158 \pm 0.010$ & $0.486 \pm 0.010$\\
		CAN-08 & $79.49751$ & $-69.74904$ & $2$ & $16.207 \pm 0.034$ & $0.498 \pm 0.040$ &
		$16.290 \pm 0.005$ & $0.274 \pm 0.007$ & $0.173 \pm 0.010$ & $0.955 \pm 0.010$\\
		CAN-09 & $79.50509$ & $-69.63110$ & $3$ & $16.472 \pm 0.026$ & $0.772 \pm 0.031$ &
		$16.560 \pm 0.005$ & $0.456 \pm 0.007$ & $0.239 \pm 0.011$ & $0.567 \pm 0.012$\\
		CAN-10 & $79.55121$ & $-69.72603$ & $2$ & $15.040 \pm 0.008$ & $0.583 \pm 0.012$ &
		$15.100 \pm 0.003$ & $0.302 \pm 0.004$ & $0.135 \pm 0.005$ & $1.097 \pm 0.005$\\
		CAN-11 & $79.56479$ & $-69.65368$ & $3$ & $15.400 \pm 0.015$ & $0.744 \pm 0.022$ &
		$15.469 \pm 0.003$ & $0.350 \pm 0.004$ & $0.250 \pm 0.007$ & $0.941 \pm 0.008$\\
		CAN-12 & $79.56815$ & $-69.65643$ & $3$ & $17.401 \pm 0.024$ & $0.556 \pm 0.039$ &
		$17.446 \pm 0.007$ & $0.292 \pm 0.010$ & $0.180 \pm 0.015$ & $0.909 \pm 0.016$\\
		CAN-13 & $79.63653$ & $-69.63896$ & $3$ & $16.378 \pm 0.019$ & $0.784 \pm 0.024$ &
		$16.463 \pm 0.004$ & $0.462 \pm 0.006$ & $0.270 \pm 0.010$ & $0.577 \pm 0.011$\\
		CAN-14 & $79.69809$ & $-69.65521$ & $3$ & $15.014 \pm 0.011$ & $0.873 \pm 0.014$ &
		$15.082 \pm 0.003$ & $0.542 \pm 0.004$ & $0.379 \pm 0.008$ & $0.533 \pm 0.009$\\
		CAN-15 & $82.09185$ & $-69.86350$ & $4$ & $17.337 \pm 0.031$ & $0.722 \pm 0.034$ &
		$17.407 \pm 0.006$ & $0.408 \pm 0.009$ & $0.170 \pm 0.015$ & $0.660 \pm 0.016$\\
		CAN-16 & $82.10721$ & $-69.82555$ & $4$ & $16.500 \pm 0.017$ & $0.616 \pm 0.024$ &
		$16.557 \pm 0.004$ & $0.291 \pm 0.006$ & $0.195 \pm 0.010$ & $0.881 \pm 0.010$\\
		CAN-17 & $82.11237$ & $-69.87718$ & $4$ & $15.002 \pm 0.010$ & $0.776 \pm 0.014$ &
		$15.035 \pm 0.002$ & $0.409 \pm 0.003$ & $0.315 \pm 0.005$ & $0.689 \pm 0.005$\\
		CAN-18 & $82.14838$ & $-69.84965$ & $4$ & $16.889 \pm 0.026$ & $0.625 \pm 0.032$ &
		$16.963 \pm 0.005$ & $0.279 \pm 0.008$ & $0.099 \pm 0.014$ & $0.908 \pm 0.012$\\
		CAN-19 & $82.27860$ & $-69.84849$ & $4$ & $17.582 \pm 0.072$ & $0.625 \pm 0.080$ &
		$17.639 \pm 0.009$ & $0.311 \pm 0.013$ & $0.222 \pm 0.021$ & $0.776 \pm 0.020$\\
		\hline
	\end{tabular}
\end{table*}


\begin{table*}
	\centering
	\caption{Cross-match of selected candidates for non-pulsating stars in the LMC classical IS with
	OGLE catalogs: ID~- name of the candidate; OGLE-II, OGLE-III, OGLE-IV~- name of the candidate in
	OGLE catalogs.}
	\label{tab:candidates}
	\begin{tabular}{llll} 
		\hline
		ID & OGLE-II & OGLE-III & OGLE-IV\\
		\hline
		CAN-01 & LMC\_SC8 224920 & LMC100.7.17275  & LMC503.13.120132\\ 
		CAN-02 & LMC\_SC8 225288 & LMC100.7.17672  & LMC503.13.120383\\ 
		CAN-03 & LMC\_SC8 312232 & LMC100.7.17331 & LMC503.13.36068 \\ 
		CAN-04 & LMC\_SC7 22149  & LMC103.5.30585  & LMC503.05.44074\\ 
		CAN-05 & LMC\_SC7 38882  & LMC103.5.92208  & LMC503.13.248\\ 
		CAN-06 & LMC\_SC7 14125 & LMC103.6.78881 & - \\ 
		CAN-07 & LMC\_SC7 14162  & LMC103.6.78891  & LMC503.05.26852\\ 
		CAN-08 & LMC\_SC7 14127  & LMC103.6.78900  & LMC503.05.26808\\ 
		CAN-09 & LMC\_SC7 30218 & LMC103.5.30386 & LMC503.05.43806 \\ 
		CAN-10 & LMC\_SC7 134320 & LMC103.6.78723 & - \\ 
		CAN-11 & LMC\_SC7 142141 & LMC103.5.30256  & LMC503.04.109410 \\ 
		CAN-12 & LMC\_SC7 142382 & LMC103.5.47427  & LMC503.04.109830\\ 
		CAN-13 & LMC\_SC7 142250 & LMC103.5.47327  & LMC503.04.109565\\ 
		CAN-14 & LMC\_SC7 142103 & LMC103.5.47191  & LMC503.04.109345\\ 
		CAN-15 & LMC\_SC3 162486 & LMC162.2.117682 & - \\ 
		CAN-16 & LMC\_SC3 170350 & LMC162.3.63638  & LMC516.23.68773 \\ 
		CAN-17 & LMC\_SC3 162131 & LMC162.2.136551 & LMC516.23.54100\\ 
		CAN-18 & LMC\_SC3 170298 & LMC162.2.136736 & LMC516.23.68723 \\ 
		CAN-19 & LMC\_SC3 282212 & LMC169.7.83593  & LMC516.23.68910 \\ 
		\hline
	\end{tabular}
\end{table*}

\section{Conclusions}
\label{sec:concl}

Based on OGLE photometric maps combined with Str\"omgren photometry and Gaia DR2 data for four fields
in the LMC we found $19$ candidates for non-pulsating stars located in empirical instability strip.
An analysis of their light curves downloaded from OGLE surveys confirmed that they are stable at the
level of a~few of mmag. Our results show that between about $21\%$ to $30\%$ of LMC giants located in
the IS might not pulsate. More observations will be required for further investigation of the
candidates, in order to find any signs of variability indicating kinds of behavior other than classical
pulsations, e.g., nonradial pulsation or binarity (where both components of the system lie in fact
outside the IS).


The relation $(u-b) - \log{P}$ plotted for classical Cepheids identified in the fields gave us two
potential candidates for binary Cepheids with hot companions and two other are questionable. These
systems might be interesting for further follow up spectroscopic observations to confirm these
suspicions.

In this work, we showed that the number of potential non-pulsating stars in the classical IS of
Cepheids is a~non-zero number and its upper limit suggests that it might be even significant
percentage of giants in the IS. The follow up spectroscopic studies shall revise this conclusion,
but basing on photometry we show it is very likely that non-pulsating stars exist in the IS. The
great potential of multiband photometry combined with spectroscopy and Gaia proper motions could
provide information about the nature of these mysterious objects.

\section*{Acknowledgements}

We thank anonymous reviewer for comments which helped to improve original manuscript.

W.N. thanks prof. Andrzej So\l{}tan for a fruitful discussion about statistics.

We thank the staff of Cerro Tololo Observatory for a~support during the observations.

W.N., G.P., Z.K., R.S. and M.G. were partly supported by the grant MAESTRO UMO-2017/26/A/ST9/00446 from
the Polish National Science Center.

B.P. acknowledges financial support for this work from the Polish National Science Center
grant SONATA 2014/15/D/ST9/02248.

Authors were also partially supported by the grant IDEAS PLUS IdP II 2015 0002 64 of the
Polish Ministry of Science and Higher Education.

The OGLE project received funding from the Polish National Science Centre grant MAESTRO no.
2014/14/A/ST9/00121

The research leading to these results has received funding from the European Research
Council (ERC) under the European Unions Horizon 2020 research and innovation program
(grant agreement No. 695099). W.N., W.G., M.G. and D.G. gratefully acknowledge financial
support for this work from the Millennium Institute of Astrophysics (MAS) of the Iniciativa
Cientifica Milenio del Ministerio de Economica, Fomento y Turismo de Chile, project IC120009.
We (W.G., G.P., and D.G.) also very gratefully acknowledge financial support for this work
from the BASAL Centro de Astrofisica y Tecnologias Afines (CATA) AFB-170002.

This work has made use of data from the European Space Agency (ESA) mission {\it Gaia}
(\url{https://www.cosmos.esa.int/gaia}), processed by the {\it Gaia} Data Processing and
Analysis Consortium (DPAC, \url{https://www.cosmos.esa.int/web/gaia/dpac/consortium}). Funding
for the DPAC has been provided by national institutions, in particular the institutions
participating in the {\it Gaia} Multilateral Agreement.








\appendix


\bsp	
\label{lastpage}
\end{document}